\documentclass[10pt,a4paper,twocolumn]{article}

\begin{document}

\title{An advance report on particle invariance in particle physics}
\author{{\small H. Y. Cui} \\
{\small Department of Physics, Beihang University}\\
{\small Beijing, 100083, China, E-mail: hycui@buaa.edu.cn}}
\date{{\small \today}}
\maketitle

\begin{abstract}
{\small \ Since particle such as molecule, atom and nucleus are composite
particle, it is important to recognize that physics must be invariant for
both the composite particle and its constituent particles, this requirement
is called particle invariance. But difficulties arise immediately because
for fermion we use the Dirac equation, for boson we use the Klein-Gordon
equation. Therefore, the particle invariance demands there is a general wave
equation for describing particle motion regardless particle class. In this
paper, three advances in this subject are reported: (1)
momentum-wavefunction relation is a general relation shared by both fermion
and boson, meets the requirement of the particle invaiance. As a test, the
momentum-wavefunction relation was directly applied to hydrogen atom, and
get the correct fine structure and spin effect for the electron. (2) the
Dirac equation and Klein-Gordon equation can be derived out from the
momentum-wavefunction relation when we abandon some higher order terms. (3)
according to the momentum-wavefunction relation a path integral method was
developed , differing from Feynman's path integral, it simplfies quantum
computation. \newline
\newline
}
\end{abstract}

\section{Introduction}

Consider a particle of rest mass $m$ and charge $q$ moving in an inertial
frame of reference with relativistic 4-vector velocity $u_\mu $, it satisfies%
\cite{Harris}

\begin{equation}
u_\mu u_\mu =-c^2  \label{s1}
\end{equation}
where there is not distinction between covariant and contravariant
components in the Cartesian coordinate system. Let $A_\mu $ denote vector
potential of electromagnetic field, substituting the momentum-wavefunction
relation

\begin{equation}
mu_{\mu }=\frac{1}{\psi }(-i\hbar \partial _{\mu }-qA_{\mu })\psi  \label{s2}
\end{equation}
into Eq.(\ref{s1}), we obtain a new quantum wave equation with single
component wavefunction

\begin{equation}
\lbrack (-i\hbar \partial _\mu -qA_\mu )\psi ][(-i\hbar \partial _\mu
-qA_\mu )\psi ]=-m^2c^2\psi ^2  \label{s3}
\end{equation}
where we regard the momentum as momentum itself but not momentum operator,
please note that Eq.(\ref{s3}) is not the Klein-Gordon wave equation. In the
recent years, H Y Cui has studied this equation for many years, it was found
that by solving Eq.(\ref{s3}) for hydrogen atom, the fine structure of
hydrogen energy can be calculated correctly, while its wavefunction has only
single component in contrast with Dirac's wavefunction, the spin effect of
electron is also revealed by Eq.(\ref{s3}) when the hydrogen atom is in a
magnetic field\cite{CuiHydr}\cite{CuiBeijing}\cite{Cuisc}. It was also found
that the Dirac wave equation and Klein-Cordon wave equation can be derived
out from Eq.(\ref{s3}) when we abandon some higher order terms or nonlinear
terms\cite{CuiDirac}. These results are easily understood regarding that
Eq.( \ref{s1}) is a general relation shared by both fermion and boson, or
any other particle.

In the present paper, we mainly report that a path integral method based on
the momentum-wavefunction relation was developed, it differs from Feynman's
path integral; as a test, the spin effect of electron are calculated by
using the path integral method, giving out correct results. It was shown
that the path integral method is a rapid quantum computation method which
satisfies the particle invariance.

\section{Path integral method}

Consider Eq.(\ref{s2}), its path integral form is given by 
\begin{equation}
\psi =e^{\frac i\hbar \int (p_\mu +qA_\mu )dx_\mu }  \label{ia1}
\end{equation}
where $p_\mu =mu_\mu $ is the momentum of the particle. In the Cartesian
coordinate system $(x_1,x_2,x_3,x_4=ict)$, from the Eq.(\ref{s1}), the
momentum components satisfy

\begin{equation}
p_1^2+p_2^2+p_3^2+p_4^2=-m^2c^2  \label{ia3}
\end{equation}
In the next sections, we test the path integral for some physical systems,
to note that it differs from Feynman's path integral.

\section{Fine structure}

In the following, we use Gaussian units, and use $m_e$ to denote the rest
mass of electron, thus

\begin{equation}
\psi =e^{\frac i\hbar \int (p_\mu +qA_\mu /c)dx_\mu }  \label{ia4}
\end{equation}

In a spherical polar coordinate system $(r,\theta ,\varphi ,ict)$, the
nucleus of hydrogen atom provides a spherically symmetric potential $%
V(r)=e/r $ for the electron $(q=-e)$, the displacement elements and vector
potential are given by 
\begin{eqnarray}
dx_r &=&dr \\
dx_\theta &=&rd\theta \\
dx_\varphi &=&r\sin \theta d\varphi \\
A_r &=&A_\theta =A_\varphi =0 \\
A_4 &=&iV=ie/r
\end{eqnarray}
Then, the wavefunction is given by

\begin{eqnarray}
\psi &=&e^{\frac i\hbar \int p_rdx_r}e^{\frac i\hbar \int p_\theta dx_\theta
}e^{\frac i\hbar \int p_\varphi dx_\varphi }e^{\frac i\hbar \int
(p_4+qA_4/c)dx_4}  \nonumber \\
&&
\end{eqnarray}
For separating the variables so that $\psi =R(r)X(\theta )\phi (\varphi
)e^{-iEt/\hbar }$ for energy eigenstate, we expect

\begin{eqnarray}
\phi (\varphi ) &=&e^{\frac i\hbar \int p_\varphi dx_\varphi }  \label{ia6}
\\
X(\theta ) &=&e^{\frac i\hbar \int p_\theta dx_\theta }  \label{ia7} \\
R(r) &=&e^{\frac i\hbar \int p_rdx_r}  \label{ia8} \\
e^{-iEt/\hbar } &=&e^{\frac i\hbar \int_0^t(p_4+qA_4/c)dx_4}  \label{ia9}
\end{eqnarray}
The angular momentum magnitude and its $z$-axis component magnitude are
denoted by $J$ and $J_z$ respectively, we have

\begin{eqnarray}
p_\varphi r\sin \theta &=&J_z\qquad (const.) \\
(\sqrt{p_\theta ^2+p_\varphi ^2})r &=&J\qquad (const.)
\end{eqnarray}
and 
\begin{eqnarray}
p_4 &=&\frac{-E-icqA_4/c}{ic}=\frac{-E-e^2/r}{ic} \\
p_r &=&\pm \sqrt{-m_e^2c^2-p_\theta ^2-p_\varphi ^2-p_4^2} \\
&=&\pm \sqrt{-m_e^2c^2-\frac{J^2}{r^2}+\frac 1{c^2}(E+\frac{e^2}r)^2}
\end{eqnarray}
thus we have 
\begin{eqnarray}
\phi (\varphi ) &=&e^{\frac i\hbar \int p_\varphi dx_\varphi }=C_1e^{\frac
i\hbar J_z\varphi }  \label{ia10} \\
X(\theta ) &=&e^{\frac i\hbar \int p_\theta dx_\theta }=C_2e^{\pm \frac
i\hbar \int_0^\theta \sqrt{J^2-\frac{J_z^2}{\sin ^2\theta }}d\theta }
\label{ia11} \\
R(r) &=&C_3e^{\pm \frac i\hbar \int_0^r\sqrt{-m_e^2c^2-\frac{J^2}{r^2}+\frac
1{c^2}(E+\frac{e^2}r)^2}dr}  \label{ia12}
\end{eqnarray}
where $C_1$, $C_2$ and $C_3$ are integral constants. Since $\phi (\varphi )$
and $X(\theta )$ must be periodic functions, and the radical wavefunction $%
R(r)$ forms a ''standing wave''\ in the range from $r=0$ to $r=\infty $ ,
these requirements demand

\begin{eqnarray}
\frac 1\hbar \int_0^{2\pi }J_zd\varphi  &=&2\pi m  \label{ia13} \\
(m &=&0,\pm 1,\pm 2...)  \nonumber \\
\frac 1\hbar \int_0^{2\pi }\sqrt{J^2-\frac{J_z^2}{\sin ^2\theta }}d\theta 
&=&2\pi k  \label{ia14} \\
(k &=&0,1,2...)  \nonumber \\
\frac 1\hbar \int_0^\infty \sqrt{-m_e^2c^2-\frac{J^2}{r^2}+\frac 1{c^2}(E+%
\frac{e^2}r)^2}dr &=&\pi s  \label{ia15} \\
(s &=&0,1,2...)  \nonumber
\end{eqnarray}
These definite integrals have been evaluated in the author's previous paper 
\cite{CuiHydr} by using the residue theorem and contour integrations in
complex space, because the last two integrands are multiple-valued functions
when over their turning points, the results are given by

\begin{eqnarray}
J_z &=&m\hbar  \label{ia16} \\
\frac 1\hbar \int\nolimits_0^{2\pi }\sqrt{J^2-\frac{J_z^2}{\sin ^2\theta }}%
d\theta &=&2\pi (\frac J\hbar -|m|)  \label{ia17} \\
J &=&(k+|m|)\hbar =j\hbar  \label{ia21}
\end{eqnarray}
\begin{eqnarray}
&&\frac 1\hbar \int\nolimits_0^\infty \sqrt{-m_e^2c^2-\frac{J^2}{r^2}+\frac
1{c^2}(E+\frac{e^2}r)^2}dr \nonumber\\
&=&\frac{\pi E\alpha }{\sqrt{m_e^2c^4-E^2}}-\pi \sqrt{j^2-\alpha ^2}
\end{eqnarray}
\begin{equation}
\frac{\pi E\alpha }{\sqrt{m_e^2c^4-E^2}}-\pi \sqrt{j^2-\alpha ^2}=\pi s
\end{equation}
where $\alpha =e^2/\hbar c$ is known as the fine structure constant.

Form the last Eq.(\ref{ia21}) , we obtain the energy levels given by

\begin{equation}
E=m_ec^2\left[ 1+\frac{\alpha ^2}{(\sqrt{j^2-\alpha ^2}+s)^2}\right]
^{-\frac 12}  \label{s16}
\end{equation}
where $j=k+|m|$, because $j\neq 0$ in Eq.(\ref{s16}), we find $j=1,2,3...$.

The result, Eq.(\ref{s16}), is completely the same as that in the
calculation of the Dirac wave equation\cite{Schiff} for the hydrogen atom,
it is just the \textbf{fine structure of hydrogen energy}.

\section{Electronic spin}

If we put the hydrogen atom into an external uniform magnetic field $B$
which is along the $z$ axis with the vector potential $(A_r,A_\theta
,A_\varphi )=(0,0,\frac 12r\sin \theta B)$, i.e. $\mathbf{B}=B\mathbf{e}_z$,
where $\mathbf{e}_z$ is the unit vector along $z$ axis. According to Eq.(\ref
{ia1}), the energy eigenstate of the hydrogen atom is described by

\begin{eqnarray}
\psi &=&R(r)X(\theta )\phi (\varphi )e^{-iEt/\hbar } \\
\phi (\varphi ) &=&e^{\frac i\hbar \int (p_\varphi +qA_\varphi /c)dx_\varphi
} \\
X(\theta ) &=&e^{\frac i\hbar \int p_\theta dx_\theta } \\
R(r) &=&e^{\frac i\hbar \int p_rdx_r} \\
e^{-iEt/\hbar } &=&e^{\frac i\hbar \int_0^t(p_4+qA_4/c)dx_4}
\end{eqnarray}
The magnitude of the angular momentum is denoted by $J$ and its component
along $z$-axis by $J_z$, then

\begin{eqnarray}
p_\varphi r\sin \theta &=&J_z\qquad (const.) \\
(\sqrt{p_\theta ^2+p_\varphi ^2})r &=&J\qquad (const.)
\end{eqnarray}
we also have the same expressions as 
\begin{eqnarray}
p_4 &=&\frac{-E-icqA_4/c}{ic}=\frac{-E-e^2/r}{ic} \\
p_r &=&\pm \sqrt{-m_e^2c^2-p_\theta ^2-p_\varphi ^2-p_4^2} \\
&=&\pm \sqrt{-m_e^2c^2-\frac{J^2}{r^2}+\frac 1{c^2}(E+\frac{e^2}r)^2}
\end{eqnarray}
but we have 
\begin{eqnarray}
\phi (\varphi ) &=&e^{\frac i\hbar \int (p_\varphi +qA_\varphi /c)dx_\varphi
}  \nonumber \\
&=&e^{\frac i\hbar \int (p_\varphi +q\frac 1{2c}r\sin \theta B)r\sin \theta
d\varphi }  \nonumber \\
&=&C_1e^{\frac i\hbar (J_z-e\frac 1{2c}r^2\sin ^2\theta B)\varphi
}=C_1e^{im\varphi } \\
X(\theta ) &=&e^{\frac i\hbar \int p_\theta dx_\theta }=C_2e^{\pm \frac
i\hbar \int_0^\theta \sqrt{J^2-\frac{J_z^2}{\sin ^2\theta }}d\theta } 
\nonumber \\
&=&C_2e^{\pm \frac i\hbar \int_0^\theta \sqrt{J^2-(m\hbar +e\frac
1{2c}r^2\sin ^2\theta B)^2/\sin ^2\theta }d\theta }  \nonumber \\
&\simeq &C_2e^{\pm \frac i\hbar \int_0^\theta \sqrt{J^2-\frac{m^2\hbar ^2}{%
\sin ^2\theta }-\frac{m\hbar er^2B}c}d\theta } \\
R(r) &=&C_3e^{\pm \frac i\hbar \int_0^r\sqrt{-m_e^2c^2-\frac{J^2}{r^2}+\frac
1{c^2}(E+\frac{e^2}r)^2}dr}
\end{eqnarray}
where $C_1$, $C_2$ and $C_3$ are integral constants, we have neglected $%
O(B^2)$ term. Since $\phi (\varphi )$ and $X(\theta )$ must be periodic
functions, and the radical wavefunction $R(r)$ forms a ''standing wave''\ in
the range from $r=0$ to $r=\infty $ , these requirements demand

\begin{eqnarray}
\frac 1\hbar \int_0^{2\pi }(J_z-e\frac 12r^2\sin ^2\theta B)d\varphi &=&2\pi
m \\
(m &=&0,\pm 1,\pm 2...)  \nonumber \\
\frac 1\hbar \int_0^{2\pi }\sqrt{J^2-\frac{m^2\hbar ^2}{\sin ^2\theta }-%
\frac{m\hbar er^2B}c}d\theta &=&2\pi k \\
(k &=&0,1,2...)  \nonumber \\
\frac 1\hbar \int_0^\infty \sqrt{-m_e^2c^2-\frac{J^2}{r^2}+\frac 1{c^2}(E+%
\frac{e^2}r)^2}dr &=&\pi s \\
(s &=&0,1,2...)  \nonumber
\end{eqnarray}
These definite integrals can be evaluated in the same way as in the author's
previous paper \cite{CuiHydr}, given by

\begin{equation}
J_z-e\frac 12r^2\sin ^2\theta B=m\hbar
\end{equation}

\begin{eqnarray}
&&\frac 1\hbar \int\nolimits_0^{2\pi }\sqrt{J^2-\frac{m^2\hbar ^2}{\sin
^2\theta }-\frac{m\hbar er^2B}c}d\theta \\
&=&2\pi (\frac 1\hbar \sqrt{J^2-\frac{m\hbar er^2B}c}-|m|)
\end{eqnarray}
we get 
\begin{equation}
J^2-\frac{m\hbar er^2B}c=(k+|m|)^2\hbar ^2=j^2\hbar ^2  \label{ia31}
\end{equation}
we have 
\begin{eqnarray}
&&\frac 1\hbar \int\nolimits_0^\infty \sqrt{-m_e^2c^2-\frac{J^2}{r^2}+\frac
1{c^2}(E+\frac{e^2}r)^2}dr  \nonumber \\
&=&\frac 1\hbar \int\nolimits_0^\infty \sqrt{-m_e^2c^2-\frac 1{r^2}(j^2\hbar
^2+\frac{m\hbar er^2B}c)+\frac 1{c^2}(E+\frac{e^2}r)^2}dr  \nonumber \\
&=&\frac{\pi E\alpha }{\sqrt{m_e^2c^4+m\hbar ecB-E^2}}-\pi \sqrt{j^2-\alpha
^2}
\end{eqnarray}

\begin{equation}
\frac{\pi E\alpha }{\sqrt{m_e^2c^4+m\hbar ecB-E^2}}-\pi \sqrt{j^2-\alpha ^2}%
=\pi s
\end{equation}
we obtain the energy levels of hydrogen atom in the magnetic field given by

\begin{equation}
E=\sqrt{m_{e}^{2}c^{4}+mec\hbar B}\left[ 1+\frac{\alpha ^{2}}{(\sqrt{%
j^{2}-\alpha ^{2}}+s)^{2}}\right] ^{-\frac{1}{2}}  \label{d8}
\end{equation}

In the usual spectroscopic notation of quantum mechanics, four quantum
numbers: $n$, $l$, $m_l$ and $m_s$ are used to specify the state of an
electron in an atom. After the comparison, we get the relations between the
usual notation and our notation.

\begin{eqnarray}
n &=&j+s,\quad s=0,1,...;j=1,2,....  \label{d10} \\
l &=&j-1,  \label{d11} \\
\quad \max (m_{l}) &=&\max (m)-1  \label{d12}
\end{eqnarray}
We find that $j$ takes over $1,2,...,n$; for a fixed $j$ (or $l$), $m$ takes
over $-(l+1),-l,...,0,...,l,l+1$. In the present work, spin quantum number
is absent.

According to Eq.(\ref{d8}), for a fixed $(n,l)$, equivalent to $(n,j=l+1)$,
the energy level of hydrogen atom will split into $2l+3$ energy levels in
the magnetic field, given by

\begin{eqnarray}
E &=&(m_ec^2+\frac{me\hbar B}{2m_ec})\left[ 1+\frac{\alpha ^2}{(\sqrt{%
j^2-\alpha ^2}+s)^2}\right] ^{-\frac 12}  \nonumber \\
&&+O(B^2)  \label{d13}
\end{eqnarray}
Considering $m=-(l+1),-l,...,0,...,l,l+1$, this effect is equivalent to the
usual Zeeman splitting in the usual quantum mechanics given by

\begin{equation}
E=E_{nl}+\frac{(m_l\pm 1)e\hbar B}{2m_ec}  \label{d14}
\end{equation}
But our work works on it without \textbf{spin} concept, the so-called spin
effect has been revealed by Eq.(\ref{d13}) without spin concept, this result
indicates that electronic spin is a kind of orbital motion. In Stern-Gerlach
experiment, the angular momentum of ground state of hydrogen atom is
presumed to be zero according to the usual quantum mechanics, thus ones need
make use of the spin. But in the present calculation, the so-called spin has
been merged with the orbital motion of the electron.

Spin concept objects to particle invariance, bear in mind that simplicity is
always a merit for the physics.

\section{Discussion}

1. influence of nucleus motion

Since the nucleus and the electron in hydrogen atom consists an atom system
with a rest center of mass, the nucleus is moving, therefore, the nucleus
should provide a modified Coulomb' force which contains both electric field
and magnetic field for the electron. In the author's previous paper\cite
{CuiCoul}, the modified Coubomb's 4-force $f$ is expressed in the Minkowsky
space as (ref. to Fig.1 in the paper)

\begin{equation}
f=\frac{kqq^{\prime }}{c^2r^3}[(u\cdot X)u^{\prime }-(u\cdot u^{\prime })X]
\end{equation}
If we consider the nucleus influence on the electron, then the calculation
of hydrogen atom can be improved further.

2. gyrated state

To note that $J_z$ and $J$ are the magnitudes of the angular momentum vector
and its $z$-axis component, since they are constants, the direction of the
angular momentum vector is definitely uncertain, thus, there are two kinds
of state: (1) when $J_z=J$, the direction of $J$ is in the $z$-axis, the
state of electron motion is in a circular or elliptical orbit; (2) when $%
J_z\neq J$, the state should be gyrated state, i.e. $J$ vector keeps its
magnitude unchanged but varies its direction, this state can only appear in
non Coulomb's field by the nucleus or external field.

3. rapid quantum computation

Since the path integral method for quantum mechanics needs not to evaluate
quantum wave equation (2 order or nonlinear ones), it definitely is a rapid
quantum computation method. This path integral method provides a great
prospects for computer computation in some research fields such as $X\alpha $%
, ab-initio, LMTO, DV, etc.

The path integral method developed in the present paper differs essentially
from Feynman's path integral.

4. the particle invariance in particle physics

Since particles such as molecules, atoms and nuclei are composite particles,
it is important to recognize that physics must be invariant for both the
composite particles and their constituent particles, this requirement is
called particle invariance\cite{Cuihep}.

To note that the Eq.(\ref{ia1}) is suitable for any kind of particle,
therefore, it satisfies the particle invariance in particle physics.

5. the explanation of the wavefunction

The wavefunction $\psi $ we employed in the calculation for hydrogen atom
differs from the wave function in the usual quantum mechanics, because it
was found that the wavefunction $\psi $ keeps $|\psi |=1$ everywhere in the
hydrogen atom. But this kind of wavefunction can interference with each
other, for the detail discussion see the papers\cite{CuiOri}\cite{Cuihep}.

6. others

The author's homepage:(1) at my institution in Chinese: www.buaa.edu.cn; (2)
at Yahoo in English : www.geocities.com/hycui

\section{Conclusion}

Using equation

\[
(mu_\mu +qA_\mu )\psi =-i\hbar \partial _\mu \psi 
\]
and its integral solution 
\[
\psi =e^{\frac i\hbar \int (p_\mu +qA_\mu )dx_\mu } 
\]
a path integral method for calculating quantum state of a particle was
developed. The approach has a great advantage: it is a rapid computation
method, because it needs only to evaluate integration for quantum problem,
instead of solving quantum wave equation as in usual quantum mechanics.

In this paper, three advances in the particle invariance are reported: (1)
momentum-wavefunction relation is a general relation shared by both fermion
and boson, meets the requirement of the particle invariance. As an example,
the momentum-wavefunction relation was directly applied to hydrogen atom,
and get the correct fine structure and spin effect for the electron. (2) the
Dirac equation and Klein-Gordon equation can be derived out from the
momentum-wavefunction relation when we abandon some higher order terms. (3)
according to the momentum-wavefunction relation a path integral method was
developed , differing from Feynman's path integral, it simplifies quantum
computation.

The present calculation is characterized by using the usual
momentum-wavefunction relation directly, it provides an insight into the
foundations of quantum mechanics. The particle invariance is a basic
principle for particle physics.

\end{document}